\pdfoutput=1

\documentclass[a4paper]{panl}

\usepackage{cite}
\usepackage{wrapfig}
\usepackage{graphicx}
\usepackage{amssymb}
\usepackage{amsfonts}
\usepackage{amsmath}
\usepackage{longtable}
\usepackage{rotating}
\usepackage{lscape}
\usepackage{epsfig}
\usepackage{multirow}

\usepackage{xcolor}

\usepackage[percent]{overpic}

\usepackage{hyperref}
\usepackage[all]{hypcap}
\hypersetup{
    colorlinks,
    citecolor=blue,
    filecolor=blue,
    linkcolor=black,
    urlcolor=blue
}

\def\pt{p_{\rm T}}

\def 	\r2{\rho_2}

\def\vf{\varphi}

\def\sNN{\mbox{$\sqrt{s_{_{\rm NN}}}$}}

\originalTeX
\begin{document}

\title{Application of Principal Component Analysis \\
to establish proper 
basis for flow studies  
in heavy-ion collisions }
\maketitle
\authors{I.\,Altsybeev $^{a,}$\footnote{E-mail: i.altsybeev@spbu.ru}}
\from{$^{a}$\,Saint-Petersburg State University \\ Universitetskaya nab. 7/9, St. Petersburg, 199034, Russia}

\begin{abstract}
  
It is shown that  Principal Component Analysis (PCA) applied  to event-by-event single-particle distributions in  A-A collisions allows  establishing the most optimal basis for anisotropic flow studies  from  data itself,  
in contrast  to manual selection of the basis functions. 
 PCA coefficients for azimuthal particle distributions   are identical to Fourier coefficients from a conventional analysis techniques. PCA  applied in longitudinal dimension reveals optimal basis that is similar to Legendre polynomial series.
Analysis in both dimensions simultaneously allows studying the coupling of the longitudinal structure 
of events with the azimuthal anisotropy of particle emission.

\end{abstract}
\vspace*{6pt}

\noindent
PACS: 25.75.Gz; 25.75.Ld

\vspace*{-0.5cm}
\label{sec:intro}
\section*{\bf Introduction}

Principal Component Analysis (PCA)
is a method for decorrelation of  multivariate data
by finding the most optimal basis for a given problem 
and thus reducing its  dimensionality.
PCA is widely applied in industry,
in particular,
 for image compression, classification and recognition tasks \cite{eigenfaces1991},
 and many branches of science, 
 see a short overview, for example, in \cite{PCA_overview_2016}. 
It was 
suggested to apply PCA 
to heavy-ion collisions data
to bring out substructures from
two-particle azimuthal correlations  \cite{Ollitrault_2015}.

In this article, PCA is applied directly to single-particle distributions in A--A collisions, 
namely,
to azimuthal ($\vf$) distribution,
distribution in pseudorapidity ($\eta$) and to two-dimensional distribution $\eta$-$\vf$. 
Mathematically,
this means that 
we take 
distribution of particles in $M$ bins in each out of $N$ events, normalize with 
a number of particles in a given event,
subtract in each bin   an event-averaged value (in order to have zero mean in each bin)
and apply PCA to the obtained $N$$\times$$M$ matrix (PCA is most often done through the singular value decomposition).
As an output from PCA, we have a set of  orthonormal eigenvectors (${\mathbf e}_i, i=1,...,M$), each of the length $M$ itself,
which are ordered in such a way that corresponding variances 
($\sigma_i, i=1,...,M$) descend from the largest to the smallest values.
We get also coefficients $c_i^k (k=1,...,N$) of PCA decomposition so that
the particle distribution in $k$-th event (denote it as  ${\mathbf x}^{(k)}$ that is a vector with $M$ elements)
 can be written as
\begin{equation}
{\mathbf x}^{(k)} = 
\sum_{i=1}^M    c_i^{(k)} \sigma_i  {\mathbf e}_i = \sum_{i=1}^M   a_i^{(k)}  {\mathbf e}_i\, ,
\end{equation}
where in the last equality the variances are absorbed into coefficients: $a_i^{(k)} \equiv c_i^{(k)} \sigma_i$.
So, the first benefit of PCA is that the data matrix  $N$$\times$$M$  is projected on a set of eigenvectors ${\mathbf e}_i$
that are the {\it  most optimal basis} for given data. As the second  benefit, 
we can keep only the first $K$ components ($K$$<$$M$)
in order to  have a good approximation for the data. An exact value of $K$ can be understood 
after a closer look at the PCA output.  

Single particle distribution, denoted by ${\mathbf x}^{(k)}$,
can actually be  $\vf$, $\eta$ or $\eta$-$\vf$ distributions -- results of the PCA applied in all the three cases are discussed below.

\vspace{-0.2cm}
\section{\bf Application of PCA to azimuthal distributions}
\label{sec:PCA_to_single_particle_distr}

\begin{figure}[b]
\centering
\begin{overpic}[width=0.995\textwidth, trim={3.9cm 0.1cm 4.9cm 1.2cm},clip]
{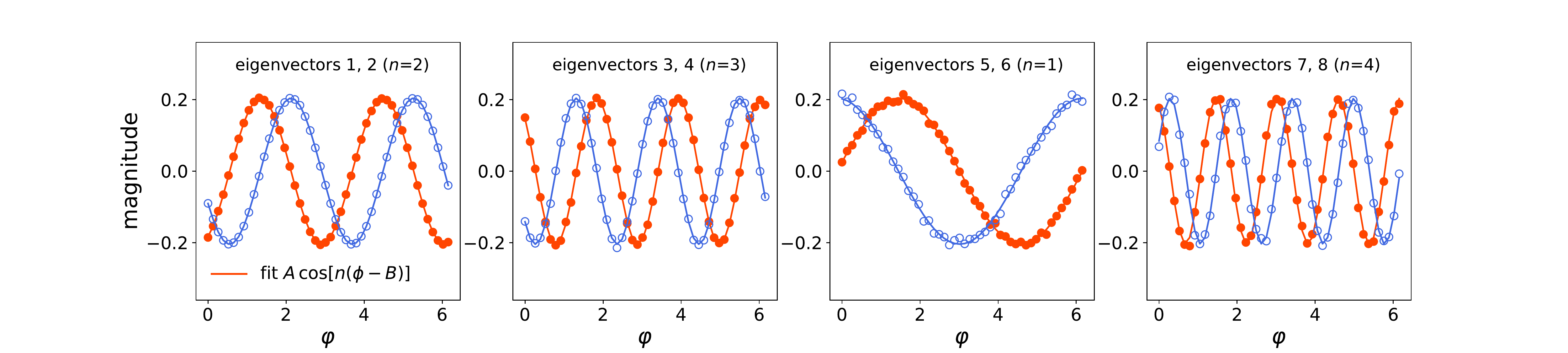} 
\put(92,5.1){ \scriptsize  \color{darkgray} AMPT  }
\end{overpic}
\vspace{-3mm}
\caption{
Eigenvectors from PCA of azimuthal distributions (48 $\vf$-bins) from AMPT events (centrality class 10-70\%).
First 8 eigenvectors are identified as Fourier harmonics of orders $n=2,3,1,4$ and are grouped in pairs
in four panels.
Ordering in $n$  reflects importance of a given harmonic (i.e. fraction of explaned variance, 
see text). 
Lines correspond to fit with a cosine function.
}
\label{eigenvectors_for_phi_48_from_AMPT}  
\end{figure}

PCA  was  applied to 1.5 mln  Pb-Pb events at $\sNN$=5 TeV
simulated in the AMPT event generator \cite{AMPT}.  
Event-by-event
$\vf$-distributions in $M$=48 bins were taken for particles within 
$|\eta| < 0.8$ and  transverse momentum range   $0.2 < \pt < 5.0$ GeV/$c$.
The first eight eigenvectors are shown in Figure \ref{eigenvectors_for_phi_48_from_AMPT},
and one may immediately notice that they 
correspond to pairs of the cosine and sine functions, i.e. the Fourier basis (with  arbitrary common phase shifts with respect to 0). 
In order to 
demonstrate this better, the eigenvectors are fitted with a cosine function 
(shown as lines) --
 the phase shift between the pairs of the functions in each panel 
  equals   $\pi$/2 with 0.01\% precision.

\begin{figure}[t]
\centering
\begin{overpic}[width=0.44\textwidth, trim={0.1cm 0.cm 1.9cm 1.75cm},clip]
{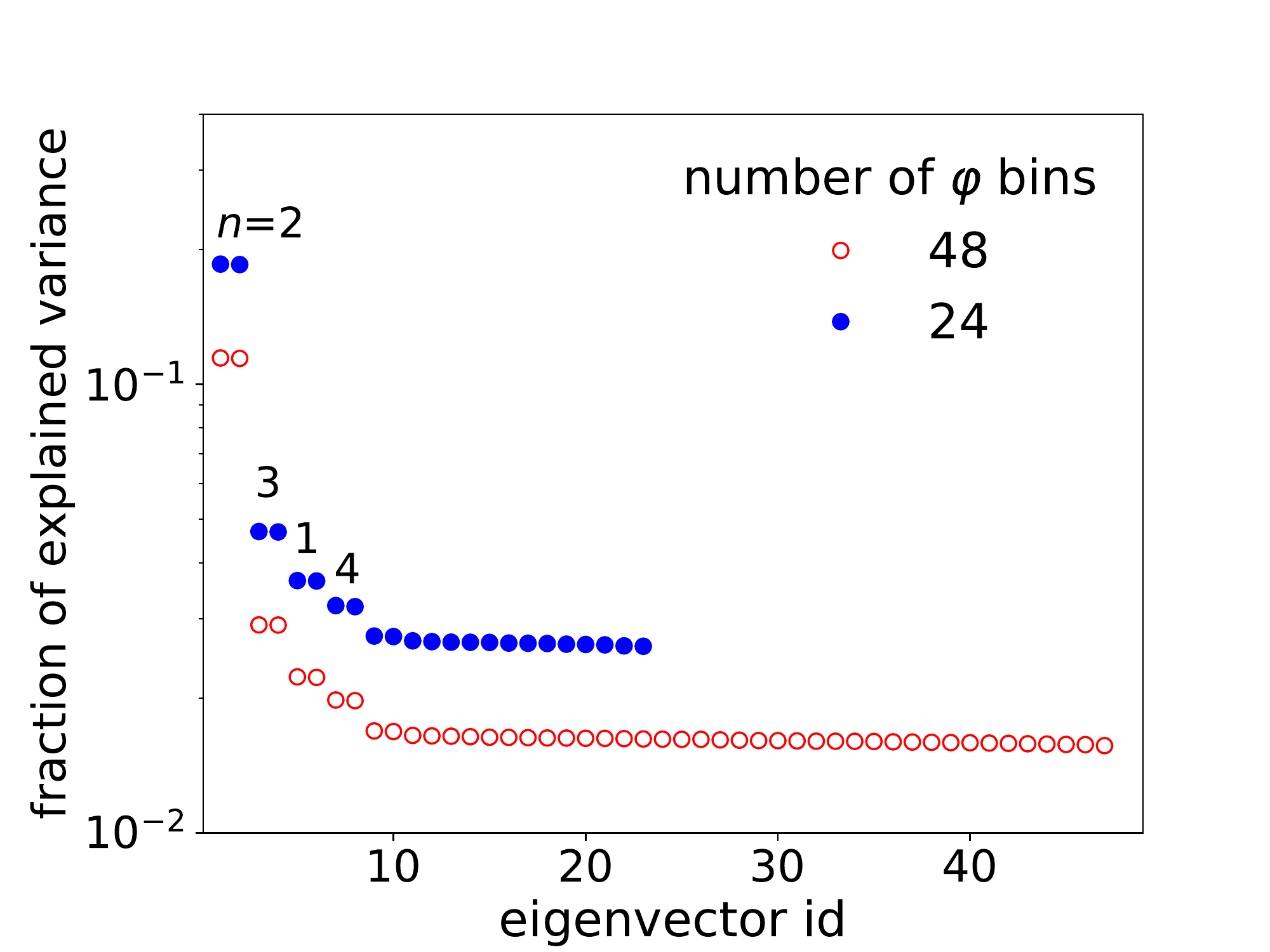} 
\put(77,31){ \footnotesize  \color{darkgray} AMPT  }
\end{overpic}\vspace{-3mm}
\caption{
Fractions of explained variances ($\sigma_i$/$\sum_{j=0}^M \sigma_j$) 
for corresponding  eigenvectors. Results are shown
for  two  numbers of $\vf$-bins:
 $M=48$ (empty markers) and 24 (full markers).
Labels $n$=2,3,1,4 for pairs of eigenvectors associated with sine and cosine functions  are placed.
}
\label{fract_of_expl_var_phiBins_24_48}
\end{figure}

Fractions of explained variances associated to 
obtained eigenvectors 
are shown in Fig.\ref{fract_of_expl_var_phiBins_24_48}. 
Pairing of variances for eigenvectors again confirms  the validity of association 
of eigenvectors with the Fourier basis. 
Eigenvectors with $i\gtrsim 10$ 
are just a statistical noise. 
It should be noted here that similar PCA analysis was performed recently in \cite{Liu_et_al:2019},
where   eigenvectors
resembles Fourier harmonics but shapes of them are somehow 
distorted\footnote{Possible explanations for such a distortion of the eigenvectors in \cite{Liu_et_al:2019} 
could be a small number of events  ($N$=$2000$) used for PCA 
or some peculiarities in event simulation process.}.

The PCA reveals the Fourier basis 
from event-by-event $\vf$-distributions 
independently of centrality class and number of bins $M$.
The explanation why PCA finds this basis as the optimal one is
in the fact that a set of  sine and cosine  functions is a natural basis for periodic or rotationally invariant problems:
 events with similar characteristic structures like elliptic flow or jets
may appear at various event plane angles, and the Fourier basis allows 
``capturing'' this information in the most optimal way.

\section{\bf Flow coefficients from PCA and correction  for statistical noise }

After the basic functions are established and interpreted,
coefficients of PCA decomposition also gain a definite meaning.
Recall that flow phenomenon in heavy-ion collisions  is usually studied using expansion
of particle azimuthal probability density in a series: 
\begin{align}
\begin{split}
\label{flow}  
f(\vf)  = \frac{1}{2\pi}
\big[ 1 + 2 \sum_{n=1}^\infty v_n  \cos\big(n(\vf-\Psi_n)\big)  \big],
\end{split}
\end{align}
where $v_n$ are  the flow 
coefficients. 
If the decomposition \eqref{flow} is applied  event-by-event,
values of $v_n$ observed in the $k$-th event
 are related to the PCA coefficients 
(associated 
to eigenvectors shown above)
as follows:
${v_2^{\rm obs}}^{(k)} =  \sqrt{M\over 2} \sqrt{ {a_1^k}^2+{a_2^k}^2}$,
${v_3^{\rm obs}}^{(k)} =  \sqrt{M\over 2} \sqrt{ {a_3^k}^2+{a_4^k}^2}$, and so on for $v_1$ and $v_4$.
 
However, since ${v_n^{\rm obs}}$ coefficients are extracted event-by-event
and a number of particles in each event is finite, 
they contain statistical fluctuations inside, 
while the task is to extract ``true'' ${v_n}$ averaged over dataset.
It can be shown that these fluctuations can be subtracted in the following way:
\begin{equation}
\label{correction_for_noise}
\big<(v_n^{\rm corr})^2\big> = \big<(v_n^{\rm obs})^2\big> - \big<(v_n^{\rm rand})^2\big> ,
\end{equation}
where $v_n^{\rm rand}$ corresponds to Fourier coefficients
extracted by applying PCA to events with randomized $\vf$-angles\footnote{
This approach is used, for instance, in \cite{Jia:2015jga} and \cite{He_Qian_Huo:2017}.}. 
In case of small flow fluctuations
and absence of non-flow effects,
the true $v_n$ can thus be estimated as $\sqrt{\big<(v_n^{\rm corr})^2\big>}$.

\begin{figure}[!b]
\centering
\begin{overpic}[width=0.67\textwidth, trim={1.58cm 1.45cm 2.65cm 2.92cm},clip]
{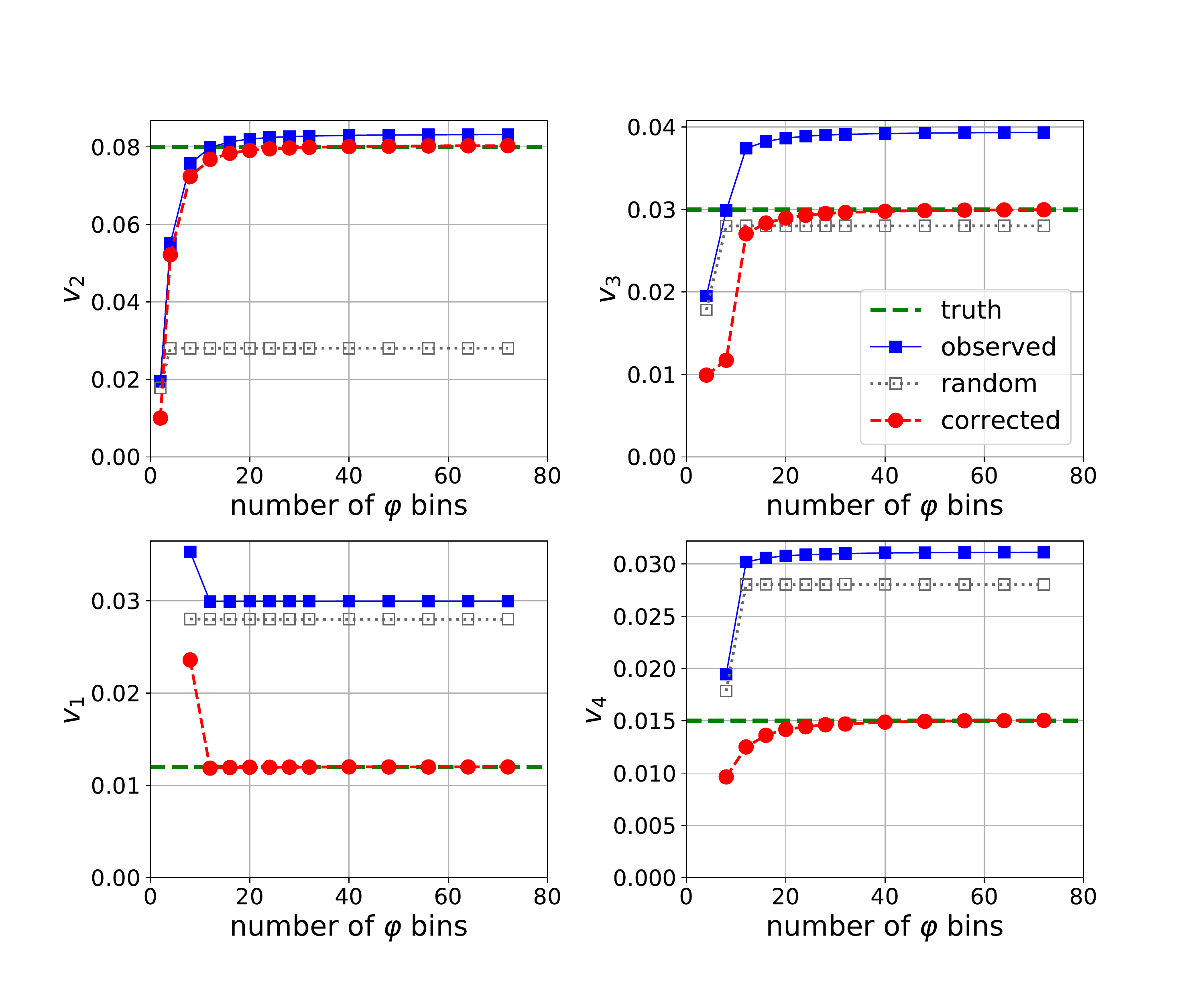} 
\put(24,71){\small \bf \color{gray} MC model }
\end{overpic}\vspace{-3mm}
\caption{
Values of $v_n$ ($n$=2,3,1,4)
before (solid blue squares) and after correction  for statistical noise (red circles)
as a function of number of $\vf$-bins used in PCA (500$k$ toy events are used in each case).
True $v_n$  are denoted by horizontal dashed lines.
Open squares show $v_n$ in  events with randomized  $\eta$ and $\vf$ of tracks. 
Number of  particles is distributed by Gauss ($\mu$=1000, $\sigma$=40).
}
\label{vn_obs_rand_corrected}
\end{figure}

Performance of the correction procedure \eqref{correction_for_noise}
is tested using a toy model with flow, where particles are distributed according to 
\eqref{flow} 
with some ``typical'' values of $v_n$. 
Analysis is done with different number of $\vf$ bins,
results are shown in Fig. \ref{vn_obs_rand_corrected}.
Different $\vf$-binnings allow one to investigate 
when PCA results become reliable 
for various harmonic orders $n$.
It can be seen that 
corrected values  (red circles)  stabilize at true values
at $n_{\vf}\gtrsim$ 30 for $v_2$, $v_3$ and $v_4$, while the true value of $v_1$
is reached somewhat earlier  (since $v_1$ measures just an overall shift of the event in azimuthal dimension that is ``captured'' already with a very few $\vf$-bins).

\begin{figure}[!t]
\centering
\begin{overpic}[width=0.995\textwidth, trim={2.9cm 0.75cm 3.65cm 1.25cm},clip]
{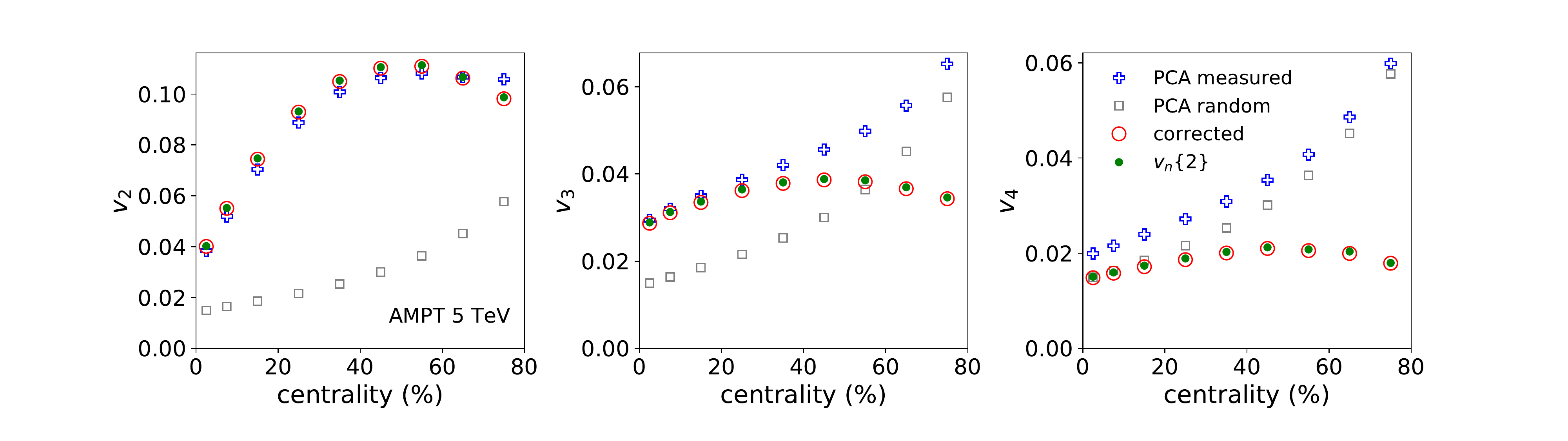} 
\end{overpic}\vspace{-3mm}
\caption{
Values of $v_n$ ($n$=2,3,4) extracted by PCA --
before correction (blue crosses) and after correction \eqref{correction_for_noise} for statistical noise (red open circles) --
as a function of centrality of Pb-Pb collisions in AMPT.
Number of  $\vf$-bins used in PCA is 48.
Open squares show $v_n$ in  events with randomized $\vf$ of tracks. 
Results obtained with the two-particle cumulant method  
are shown as green close circles.
Centrality classes are determined using percentiles of multiplicity distribution
in forward acceptance that corresponds to VZERO detector in ALICE experiment.
}
\label{v234_AMPT}
\end{figure}

In order to test robustness of $v_n$ extracted with PCA,
this analysis was applied to Pb-Pb events at $\sNN$=5 TeV
simulated in AMPT event generator
(Fig.\ref{v234_AMPT}, corrected 
PCA results   
for $v_2$, $v_3$ and $v_4$ are shown   
as open circles)
and compared to
calculations with the traditional two-particle cumulant method 
($v_n\{2\}$, full circles).
Correspondence between 
the values  justifies again the possibility to extract  $v_n$  with PCA.
It is important to note also that other conventional analyses, like symmetric cumulants
and event-plane correlations, are also possible 
 with the azimuthal PCA.

\newpage
\vspace{-0.1cm}
\section{\bf Longitudinal harmonics from PCA }
\vspace{-0.1cm}

\begin{figure}[b]
\centering
\begin{overpic}[width=0.42\textwidth, trim={0.1cm 0.05cm 1.3cm 1.6cm},clip]
{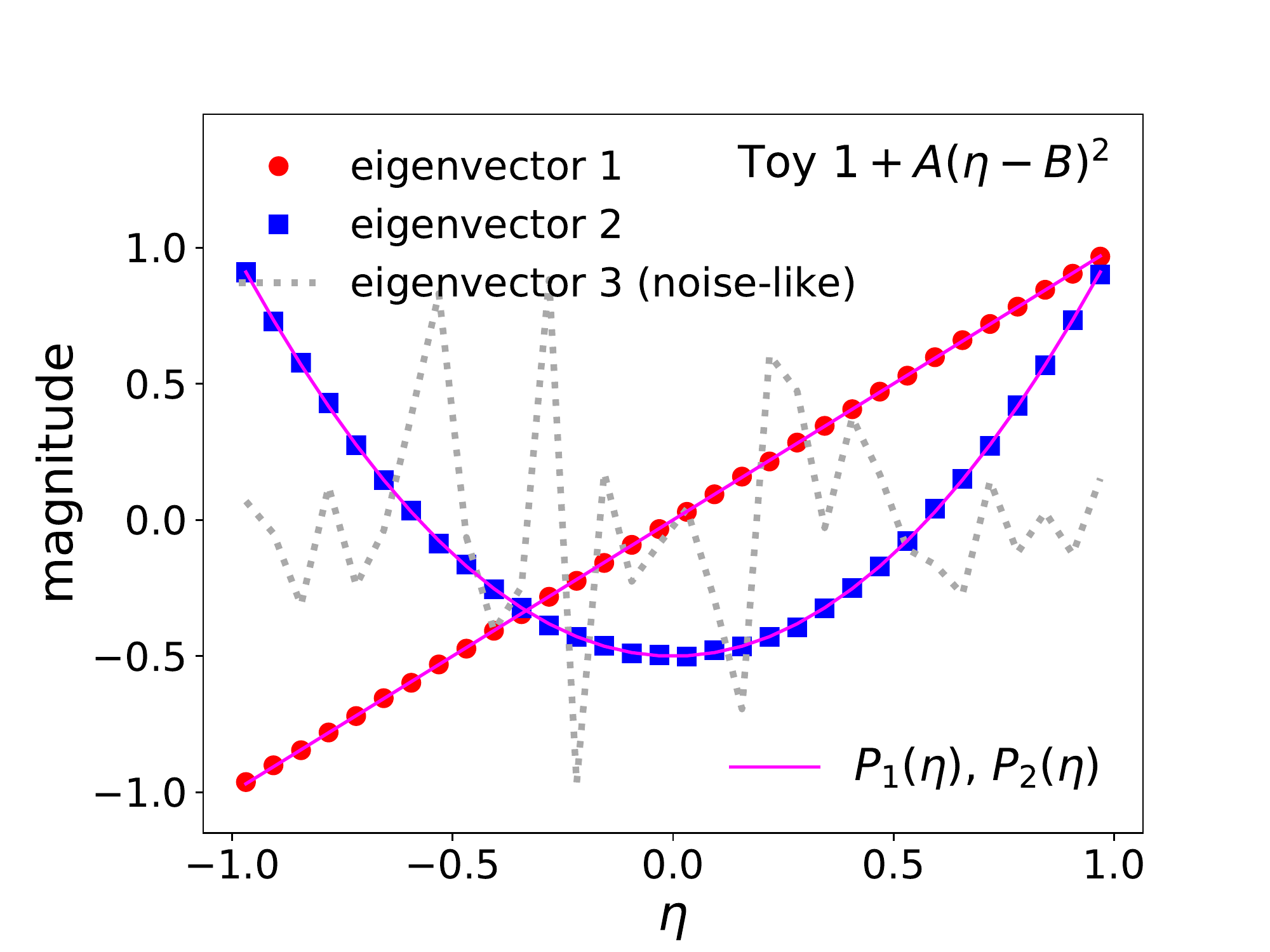} 
\end{overpic}  
\hspace{0.2cm}
\begin{overpic}[width=0.42\textwidth, trim={0.1cm 0.05cm 1.3cm 1.6cm},clip]
{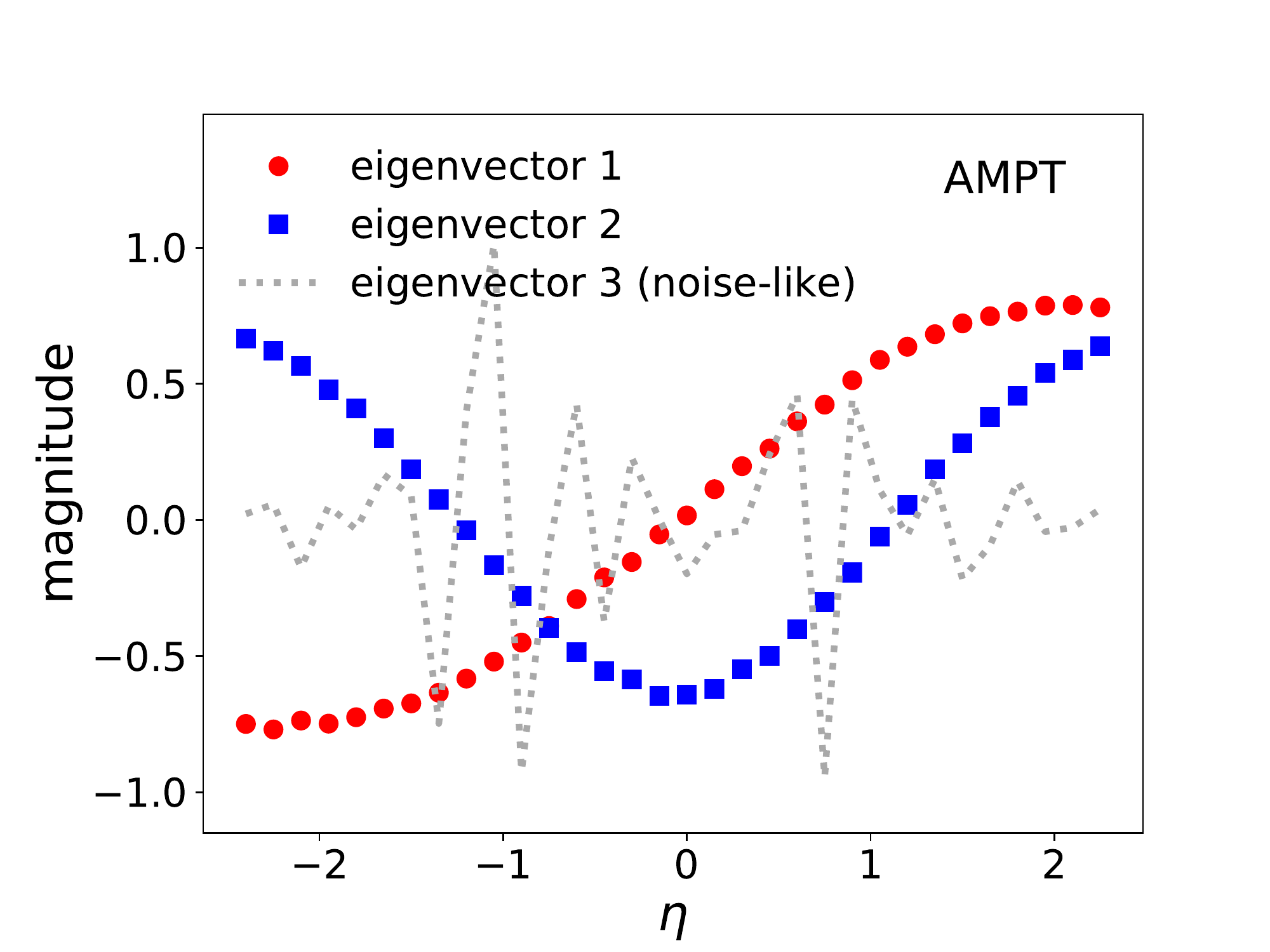} 
\end{overpic}
\vspace{-3mm}
\caption{
The first three PCA eigenvectors
obtained for the ``random parabola'' toy model (left panel)
and from AMPT for tracks with $|\eta|$<2.4 (right panel). In the left panel, the first two Legendre polynomials are drawn as lines.
 In both panels, the third eigenvectors 
are consistent with the statistical noise.
}
\label{toy_walking_parabola_components}
\end{figure}

While the Fourier basis as the best option for azimuthal distributions
was somewhat expected, it is not so obvious which basis is optimal for longitudinal ($\eta$) dimension.
It was suggested to quantify longitudinal structure of events
using Chebyshev \cite{Bzdak:2012tp} or Legendre polynomials  \cite{Jia:2015jga}
in some pseudorapidity range $[-Y ,Y ]$,  without a strong motivation for a particular choice.

The question of a proper basis for $\eta$-dimension  
can be addressed using PCA.
First of all, when does this or that polynomial basis appear in PCA?
To answer that, let us take a toy model of ``random parabola'', 
where the particle $\eta$-density in each event  is sampled according to expression  $\rho(\eta) \sim 1+A(\eta-B)^2$  with $A$ and $B$ being random numbers. It turns out that
PCA  reveals the basis of $P_1(\eta)=\eta$ and $P_2(\eta)={1\over 2} (3\eta^2-1)$, which are the first two Legendre polynomials. This is demonstrated in the left panel in Figure \ref{toy_walking_parabola_components}.

However, in a more realistic case of AMPT events 
 eigenvectors from PCA have different shapes
(right panel in Fig.\ref{toy_walking_parabola_components}).
Mathematically, this indicates that a set of these orthonormal polynomials
has its own unique weight function (recall, that  for the Legendre polynomials 
the weight function equals 1).
Moreover,  it can be shown that, unlike azimuthal case, 
PCA basis in $\eta$-dimension depends on kinematic cuts
($\eta$- and $\pt$-ranges) and physics of the collisions
(for example, results differ between AMPT and HIJING event generators).
It is interesting to get PCA eigenvectors from real A--A events
and compare with model results.

\begin{figure}[b]
\centering
\begin{overpic}[width=0.99\textwidth]
{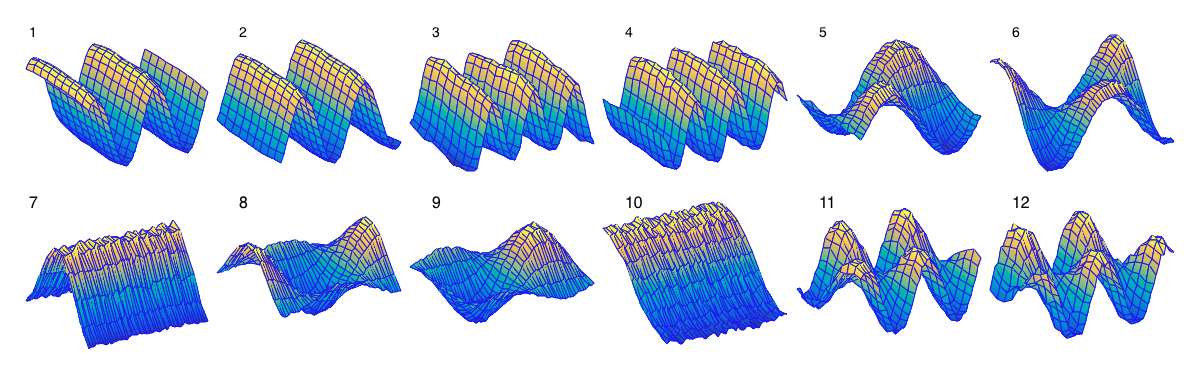} 
\end{overpic}\vspace{-4.5mm}
\caption{
First 12  eigenvectors from PCA applied to
$\eta$-$\vf$ distributions
in AMPT Pb-Pb events (centrality class 20-30\%).
Ordering is according to decrease of explained variance values.
}
\label{PCA_2D}
\end{figure}

\vspace{-0.1cm}
\section{\bf Two-dimensional case }
\vspace{-0.1cm}

Finally, the PCA can be straightforwardly 
applied for single-particle densities in two 
 dimensions, in particular, to $\eta$-$\vf$ distributions.
Eigenvectors for AMPT semi-central collisions (centrality 20-30\%)
are shown in Figure \ref{PCA_2D}
for the case of  $M$=$\eta$$\times$$\vf$=10$\times$48=480 bins.
We may note the pairs of ``azimuthal''  harmonics 
 (1,2), (3,4), etc.
that are nearly uniform in $\eta$:
it was checked that corresponding event-averaged 
$v_n$ values agree with purely azimuthal 
PCA presented above.
Longitudinal eigenvectors 7 and 10 are uniform in $\vf$,
their shapes are the same as in the right panel in Fig.\ref{toy_walking_parabola_components}.
Finally,  ``mixed'' (or ``twisted'') $\eta$-$\vf$ harmonics appear, namely,
 pairs (5,6), (8,9), (11,12). A closer look shows that
 these mixed eigenvectors can be factorized 
into $\vf$-- and $\eta$--parts 
(since PCA components must be able to capture  
different structures in $\eta$ at any azimuthal rotation).
Thus, event-by-event particle densities can be decomposed 
according to 
\begin{equation}
\rho(\eta, \vf)  = 
{1\over 2\pi} \sum_{k=0}^{K_{\varphi} }  \sum_{l=0}^{K_{\eta} }   a_{k,l}
\Phi_k (\varphi)  {\rm H}_l(\eta) \hspace{0.1cm} ,   
\end{equation}
where 
$\Phi_k (\varphi)$ denotes the azimuthal part (it can be 
written 
as 
$2\cos \big[ k(\vf-\Psi_k)\big]$),
the longitudinal part is denoted as  ${\rm H}_l(\eta)$,
 $a_{k,l}$ are the decomposition coefficients,
 $K_{\varphi}$ and $K_{\eta}$ stand for  cut-off 
numbers of harmonics to consider.
This decomposition
could be used, for instance, in studies 
of the longitudinal decorrelation of harmonic flow
as an alternative to other methods like  \cite{Jia:2014vja, BozekQM2018}.
Another possible application of this 2D-analysis  is the 
study of rapidity dependence of the directed flow.
Detailed discussion of the two-dimensional PCA
is out of scope of the present  paper.

 \newpage
\vspace{-0.3cm}
\section*{\bf Conclusion }
\vspace{-0.2cm}




Application of PCA to single-particle distributions
in A-A collisions gives a $hint$ how a proper (most optimal) basis should look like.
It was shown how PCA coefficients could be corrected 
for statistical noise.
For azimuthal dimension,   PCA confirms 
that the basis of Fourier harmonics is a proper choice, since it is natural 
for rotationally invariant problems.
In case of longitudinal dimension, 
a set of PCA 
eigenvectors  is not a “standard” one -- 
the most optimal basis of orthogonal functions
depends on given data (collision system, energy, acceptance).
 Finally, PCA was applied to two-dimensional $\eta$-$\vf$ distributions,
where ``twisted'' harmonics are revealed. This approach may be of practical use
in studies of longitudinal decorrelation of collective flow.



\section*{\bf Acknowledgements }
This study is supported  by Russian Science Foundation, grant 17-72-20045.
Author thanks  Andrey Erokhin and Evgeny Andronov for discussions
and interest in this work.



\begin{thebibliography}{1}
\def\selectlanguageifdefined#1{
\expandafter\ifx\csname date#1\endcsname\relax
\else\selectlanguage{#1}\fi}
\providecommand*{\href}[2]{{\small #2}}
\providecommand*{\url}[1]{{\small #1}}
\providecommand*{\BibUrl}[1]{\url{#1}}
\providecommand{\BibAnnote}[1]{}
\providecommand*{\BibEmph}[1]{\emph{#1}}
\ProvideTextCommandDefault{\cyrdash}{\hbox to.8em{--\hss--}}
\providecommand*{\BibDash}{\ifdim\lastskip>0pt\unskip\nobreak\hskip.2em\fi
\cyrdash\hskip.2em\ignorespaces}

\bibitem{eigenfaces1991}
\BibEmph{M. Turk and A. Pentland}~// \href{https://www.mitpressjournals.org/doi/10.1162/jocn.1991.3.1.71}{Journal of Cognitive Neuroscience}. \BibDash
\newblock 1991.   \BibDash
\newblock V. 3 (1). \BibDash
\newblock P.~71-86. 

\bibitem{PCA_overview_2016}
\BibEmph{I.T. Jolliffe and J.Cadima}~// \href{https://royalsocietypublishing.org/doi/10.1098/rsta.2015.0202}{Phil. Trans. R. Soc. A}. \BibDash
\newblock 2016.   \BibDash
\newblock V. 374. \BibDash
\newblock P.~20150202. 


\bibitem{Ollitrault_2015}
\BibEmph{R. S. Bhalerao, J.Y. Ollitrault, S. Pal and D. Teaney}~// \href{https://journals.aps.org/prl/abstract/10.1103/PhysRevLett.114.152301}{Phys. Rev. Lett.}. \BibDash
\newblock 2015.   \BibDash
\newblock V. 114. \BibDash
\newblock P.~152301. 

\bibitem{AMPT}
\BibEmph{Z.-W. Lin et al.}~// \href{https://doi.org/10.1103/PhysRevC.72.064901}{Phys. Rev. C}. \BibDash
\newblock 2005.   \BibDash
\newblock V. 72. \BibDash
\newblock P.~064901. 

\bibitem{Liu_et_al:2019}  
\BibEmph{L. Ziming, Z. Wenbin and S. Huichao}
\newblock 2019.   \BibDash
 arXiv:1903.09833 [nucl-th].

\bibitem{Jia:2015jga}  
\BibEmph{J. Jia, S. Radhakrishnan and M. Zhou}~// \href{https://journals.aps.org/prc/abstract/10.1103/PhysRevC.93.044905}{Phys. Rev. C}. \BibDash
\newblock 2016.   \BibDash
\newblock V. 93. \BibDash
\newblock P.~044905. 

 
\bibitem{He_Qian_Huo:2017}  
\BibEmph{R. He, J. Qian, L. Huo}~\BibDash
\newblock 2017 \BibDash
\newblock arXiv:1702.03137 [nucl-th].

\bibitem{Bzdak:2012tp} 
\selectlanguageifdefined{english}
\BibEmph{A. Bzdak and D. Teaney}~// \href{10.1103/PhysRevC.87.024906}{Phys. Rev. C}. \BibDash
\newblock 2013.  \BibDash
\newblock V. 87. \BibDash
\newblock P.~024906. 

\bibitem{Jia:2014vja}
\BibEmph{J. Jia and P. Huo}~// \href{https://journals.aps.org/prc/abstract/10.1103/PhysRevC.90.034905}{Phys. Rev. C}. \BibDash
\newblock 2014.  \BibDash
\newblock V. 90. \BibDash
\newblock P.~034905.  

\bibitem{BozekQM2018}
\BibEmph{P. Bozek}~// \href{https://www.sciencedirect.com/science/article/pii/S0375947418301994}{Nucl. Phys. A}. \BibDash
\newblock 2019.  \BibDash
\newblock V. 982. \BibDash
\newblock P.~335-338.  




\end{thebibliography}
\end{document}